\documentclass[a4paper,11pt]{JHEP3}
\usepackage[dvips]{graphicx}
\usepackage{ifthen}
\usepackage{amsmath}
\usepackage{epsfig}
\usepackage[errorshow]{tracefnt}

\newcommand\nn{\nonumber}

\newcommand\beal{\begin{align}}
\newcommand\eeal{\end{align}}
\newcommand\benu{\begin{enumerate}}
\newcommand\eenu{\end{enumerate}}
\newcommand\bit{\begin{itemize}}
\newcommand\eit{\end{itemize}}

\newcommand{\be}{\begin{equation}}
\newcommand{\la}{\label}
\newcommand{\ee}{\end{equation}}
\newcommand{\bd}{\begin{displaymath}}
\newcommand{\ed}{\end{displaymath}}
\newcommand{\bq}{\begin{equation}}
\newcommand{\eq}{\end{equation}}

\newcommand\ZZ{{\mathbb Z}}

\newcommand\al{\alpha}
\newcommand\ta{\tau}
\newcommand\ga{\gamma}
\newcommand\de{\delta}
\newcommand\om{\omega}

\newcommand\Om{\Omega}

\title{11D supergravity at ${\cal O}(l^3)$}

\author{Dimitrios Tsimpis\\ Institute for Theoretical Physics
\\ G\"{o}teborg University and 
\\Chalmers University of Technology
\\SE-412 96 G\"{o}teborg, Sweden}

\abstract{ We compute certain spinorial cohomology groups 
controlling possible supersymmetric 
 deformations of eleven-dimensional supergravity up to order $l^3$ in the 
Planck length. At ${\cal O}(l)$ and ${\cal O}(l^2)$ the
spinorial cohomology groups are
trivial and therefore the theory cannot be deformed 
supersymmetrically. 
At  ${\cal O}(l^3)$ the corresponding spinorial cohomology group 
is generated by
 a nontrivial element. On an eleven-dimensional
manifold $M$ such that $p_1(M)\neq 0$, this 
element corresponds to a 
supersymmetric deformation of the theory, which 
can only be redefined away at the cost of shifting 
the quantization condition of the four-form field strength.}

\preprint{G\"{o}teborg ITP preprint}
\begin{document}

\section{Introduction and summary}

In search of signatures of purely M-theoretic effects one may  
try to 
go beyond the limiting approximation of ordinary 11D supergravity, by 
including higher-order derivative (curvature) corrections. 
Despite considerable progress, however, a tractable 
microscopic definition of M-theory (see \cite{duff, t} for reviews)
on general backgrounds is still lacking 
and one would therefore have to resort to indirect computational methods.  
Supersymmetry, provided it will prove restrictive enough, 
is at present our best hope for addressing such corrections 
directly in eleven dimensions. 
The problem was analyzed within the framework of eleven-dimensional 
superspace in \cite{cgnn}, 
producing some partial results. For a review 
of the literature on $R^4$ corrections in type II string theory 
and an attempt at lifting string-theory results 
to eleven dimensions, see \cite{pvw, pvwb}.

Addressing the issue of higher-order superinvariants can be 
done systematically using spinorial-cohomology methods. 
The concept of spinorial cohomology (SC) 
was introduced in \cite{cnta} and further elaborated in \cite{cntc}. 
A purely tensorial definition was subsequently given in \cite{ht}.
SC has found a number of applications in ten-dimensional 
SYM \cite{cnta, cntb, cntd} and eleven-dimensional   
supergravity \cite{cntc, ht}.  
The ${\cal O}(l^4)$ corrections\footnote{In this paper we use  
$l$ to denote the Planck length.} to 
the worldvolume theory of the membrane in eleven dimensions 
were derived using SC in the context of the superembedding 
formalism, in \cite{hklt}. Similar methods 
were used recently in investigating higher-order corrections 
to the world-volume theory of the D9 brane \cite{kd}. 
SC with unrestricted coefficients can be shown to be
equivalent to pure-spinor \cite{hpa, hpb} cohomology, 
which was recently
used by Berkovits in the covariant quantization of the 
superstring \cite{ba, bb}. An alternative method of 
computing the cohomology by relaxing the pure-spinor 
constraint was considered in \cite{gr}.

Although  
the first supersymmetric correction 
to eleven-dimensional supergravity is expected to occur at order $l^6$, 
the existence of superinvariants already  
at lower orders in the Planck length has not been rigorously excluded. 
In this note we examine the possibility of supersymmetric 
corrections to eleven-dimensional supergravity, up to order $l^3$. 
As explained in the following,  
such corrections are controlled by certain spinorial-cohomology groups,  
which we compute to order $l^3$ in section 4. We find that up to 
order $l^2$ the relevant groups are trivial and therefore there are no
possible supersymmetric deformations of the theory. 

At $l^3$ we find that 
the corresponding spinorial-cohomology group is one-dimensional, 
therefore there exists a unique superinvariant at this order.  
On a (spin, orientable) spacetime manifold $M$ such that the 
first Pontryagin class $p_1(M)$ vanishes, this superinvariant can be 
removed by an appropriate field redefinition of the 
three-form superpotential ($C$). 
However, on a topologically nontrivial spacetime such that 
$p_1(M)\neq 0$, the superinvariant cannot be redefined away without 
changing the quantization condition of the four-form field strength 
($G=dC$, locally). 
The latter is determined, by requiring quantum consistency of the theory, 
to be of the form \cite{w}
\be
[\frac{G}{2\pi l^3}]-\frac{1}{4}p_1(M)\in H^4(M,\ZZ)~,
\ee
where the brackets denote the cohomology class.  
Recently a lot of effort has been invested in understanding the 
precise 
mathematical nature of the $C$-field on spacetime 
manifolds of nontrivial topology 
(see \cite{w,1,2,3,4,5} for an inexhaustive list 
of related literature). It would clearly be of interest to 
understand the significance of our result in relation to this issue.

In the next section we review the tensorial definition
of SC. Relevant aspects of the superspace formulation of 
ordinary eleven-dimensional supergravity are included in section 3 and in the 
appendix. Section 4 contains the computation of the 
spinorial-cohomology groups to order $l^3$.

\section{Spinorial cohomology}

In this section we give a summary of the 
tensorial definition of
spinorial cohomology for superforms (see \cite{ht} for
a more detailed discussion). 
We will suppose that the tangent bundle is a direct sum of the odd
and even bundles and that the supermanifold 
is equipped with a connection with Lorentzian
structure group. 

The space of forms admits a natural bigrading according to the
degrees and Grassmann character of the forms. Let us denote 
by $\Om^{p,q}$ the space of
superforms $\om$ with $p$ even and $q$ odd components
\be
\om_{a_1\dots a_p\al_1\dots\al_q}\in \Om^{p,q}~.
\ee
The exterior derivative $d$ has the following 
action on  $\Om^{p,q}$,
\be
d:~~\Om^{p,q}\rightarrow\Om^{p+1,q}\oplus \Om^{p,q+1} 
\oplus\Om^{p-1,q+2}\oplus\Om^{p+2,q-1}~. 
\ee
Following
\cite{bl} we split $d$ into its various components with
respect to the bigrading
 \be
 d=d_0 + d_1 + \ta_0 + \ta_1~,
 \la{3.3}
 \ee
where $d_0$ ($d_1$) is the even (odd) derivative with bidegrees
$(1,0)$ and $(0,1)$ respectively, while $\ta_0$ and $\ta_1$ have
bidegrees $(-1,2)$ and $(2,-1)$. These two latter operators are
purely algebraic and involve the dimension-zero and
dimension-three-halves components of the torsion tensor
respectively.  The fact that $d^2=0$ implies in particular
that $\ta_0^2=0$. We can therefore consider the $\ta_0$ cohomology
groups
 \be
 H^{p,q}_{\ta}=\{\om\in \Om^{p,q}|~\ta_0\om=0\}/\{ \ta_0 \lambda,
 ~\lambda\in \Om^{p+1,q-2}\}~.
 \la{3.10}
 \ee
We can now define a spinorial derivative $d_F$ which will act on
elements of $H^{p,q}_{\ta}$. For $\om\in[\om]\in H^{p,q}_{\ta}$ we set
 \be
 d_F[\om]:=[d_1 \om]~.
 \la{3.11}
 \ee
It is easy to check that this is well-defined, i.e. $d_1\om$ is
$\ta_0$-closed, and $d_F [\om]$ is independent of the choice of
representative. Moreover it is straightforward 
to show that $d_F$ is nilpotent. 
The spinorial cohomology groups are defined as
\begin{align}
H^{p,q}_F&:=H^{p,q}(d_F|H_{\ta})\nn\\
&:=\{\om\in H_\tau^{p,q}|~d_F\om=0\}/\{ d_F \lambda,
 ~\lambda\in H_\tau^{p+1,q-2}\}~.
\end{align}
If we are interested in deformations of the theory, 
we need to consider the
above cohomology groups with coefficients restricted to
be tensorial functions of the physical fields 
of the theory and their derivatives. 
We will denote these groups by $H^{p,q}_F(phys)$.

\section{Undeformed 11D supergravity}

Eleven-dimensional supergravity \cite{cjs} 
admits a superspace formulation \cite{cf, brh}. 
Let $A=(a,\al)$; $a=0\dots 10, ~\al=1\dots 32$, be a flat superspace index
and let $E^A=(E^a,E^\al)$ be the coframes of the $(11|32)$ supermanifold.
Moreover, let us introduce a connection one-form $\Om_{A}{}^B$ with
Lorentzian structure group.
The supertorsion and supercurvature are given by
\begin{align}
T^A &=DE^A:=dE^A + E^B \Om_B{}^A=\frac{1}{2} E^C E^B T_{BC}{}^A \nn\\
R_A{}^B &=d\Om_A{}^B +\Om_A{}^C\Om_C{}^B=\frac{1}{2} E^D E^C R_{CD,A}{}^B
\la{2.1}
\end{align}
and obey the Bianchi identities 
\begin{align}
DT^A&=E^B R_B{}^A \nn\\
DR_B{}^A&=0~.
\la{parker}
\end{align}
Note that for a Lorentzian structure group the second  Bianchi identity 
follows from the first \cite{dragon}.
In a purely geometrical
definition in terms of the supertorsion, it was shown
in \cite{hw} that the equations of motion of 
11D supergravity follow from the constraint
\be
T_{\al\beta}{}^{a}= -i(\ga^a)_{\al\beta}~.
\la{plato}
\ee
In this formulation the physical fields of the theory, 
the graviton, the gravitino and the 
three-form potential, appear through their covariant field strengths. 
Namely, the curvature $R_{abc}{}^d$ is identified with the
top component of the supercurvature, the gravitino field-strength
$T_{ab}{}^\al$ 
is identified with the dimension three-halves component of the
supertorsion, while the four-form field strength $G_{abcd}$ 
appears in the dimension-one components 
of the supertorsion and supercurvature. 
For completeness we have included in the appendix
all nonzero components of the supertorsion and the supercurvature, the
action of the spinorial derivative on the physical 
field strengths and their equations of motion.

The theory admits an alternative 
formulation in terms of a closed superfour-form 
$G_{ABCD}$,
\be
D_{[A}G_{BCDE\}}+2T_{[AB|}{}^FG_{F|CDE\}}=0~.
\ee
%
%
In this description, undeformed supergravity is recovered 
by imposing the constraint that the lowest component
of the superfour-form vanishes \cite{cgnn, nr}, 
\be
G_{\al\beta\ga\de}=0~.
\ee

\section{Deformations}

It was pointed out in \cite{cntc, ht}
that from the point of view of the 
superfour-form formulation of supergravity, the physical fields of the theory  
are elements of $H^{0,3}_F$ while supersymmetric 
deformations are parameterized
by elements $G_{\al\beta\ga\de}$ such that
\be
G_{\al\beta\ga\de} \in H_F^{0,4}(phys) ~.
\label{coltrane}
\ee
The content of 
equation (\ref{coltrane}) can be restated explicitly as follows \cite{ht}: 
Supersymmetric deformations of the theory are parameterized 
by objects $G_{\al\beta\ga\de}$ such that
\begin{alignat}{3}
G_{\al\beta\ga\de} &={1\over
8}(\gamma^{a_1a_2})_{(\al\beta}
(\gamma^{b_1b_2})_{\ga\de)}
A_{a_1a_2;b_1b_2}  &\qquad  \nn\\
&+{1\over 240}(\gamma^{a_1\dots a_5})_{(\al\beta}
(\gamma^{b_1b_2})_{\ga\de)}
B_{a_1\dots a_5;b_1b_2}  & \nn\\
&+{1\over 28800}(\gamma^{a_1\dots a_5})_{(\al\beta}
(\gamma^{b_1\dots b_5})_{\ga\de)} C_{a_1\dots
a_5;b_1\dots b_5}  & ~,
\label{rollins}
\end{alignat}
where $A,~B,~C$ are irreducible $(p,q)$-tensors satisfying 
\begin{align}
I_A:=D_\al \left(A-\frac{7}{5}B\right)\vert_{(02001)} &=0\nn\\
I_B:=D_\al \left(B-\frac{3}{10}C\right)\vert_{(01003)} &=0\nn\\ 
I_C:=D_\al C\vert_{(00005)} &= 0~.
\label{shorter}
\end{align}
The vertical bars in (\ref{shorter}) denote projection onto the 
representations indicated by the corresponding Dynkin weights \footnote{ 
Explicit expressions for the projections and an explanation 
of the representation-theoretical notation used here can be found
in \cite{ht}, to which the reader is referred for more details.}. 
Moreover, $G_{\al\beta\ga\de}$ is defined 
modulo shifts of the form
\be
G_{\al\beta\ga\de} \rightarrow G_{\al\beta\ga\de} + 
D_{(\al}C_{\beta\ga\de)}~,
\la{brecker}
\ee
where 
\be
C_{\al\beta\ga}=(\ga^{ab})_{(\al\beta|}V_{ab|\ga)}
+(\ga^{a_1\dots a_5})_{(\al\beta|}U_{a_1\dots a_5|\ga)}~
\la{jarret}
\ee
and $V_{ab\al}$, $U_{a_1\dots a_5\al}$ are irreducible 
(gamma-traceless) tensor-spinors. All fields in (\ref{rollins}), 
(\ref{jarret}) 
are to be understood as tensorial functions of the physical
field-strengths of the theory and their derivatives.

Note that once a $G_{\al\beta\ga\de}\in H_F^{0,4}(phys)$  has been
determined, this information can be fed into the 
Bianchi identities 
in order to derive 
the equations of motion, and therefore the Lagrangian, 
of the deformed theory.

\subsection{Deformations at ${\cal O}(l)$, ${\cal O}(l^2)$ 
and ${\cal O}(l^3)$}

The canonical dimensions of the physical field-strengths of the theory 
are as follows
\begin{align}
[G_{abcd}]&=l^{-1}\nn\\
[T_{ab}{}^\al]&=l^{-3/2}\nn\\
[R_{abcd}]&=l^{-2}~.
\end{align}
The dimensions of the fields in (\ref{rollins}), 
(\ref{jarret}) are
\begin{align}
[A]=[B]=[C]&=l\nn\\
[V]=[U]&=l^{3/2}~.
\end{align}
At any given order ${\cal O}(l^n)$, $n\geq 0$, these fields should
be expressed as functions $l^n f(G,T,R)$ 
of the physical field-strengths, where $f$ does not depend on $l$.

Clearly, at order ${\cal O}(l)$ there can be no fields $A$, $B$, $C$ 
satisfying these requirements. At order ${\cal O}(l^2)$ the combination
$l^2 G_{abcd}$ has the same dimension as $A$, $B$, $C$, but  
transforms in the `wrong' representation. We therefore conclude
that at ${\cal O}(l)$, ${\cal O}(l^2)$ the group $H_F^{0,4}(phys)$ 
is trivial and the theory does not admit
 supersymmetric deformations.

At order ${\cal O}(l^3)$ dimensional analysis 
and representation theory 
reveals that the most general expression for the fields 
$A$, $B$, $C$ is
\begin{align}
A_{a_1a_2,b_1b_2}&
=l^3\left( c_1~G_{a_1a_2}{}^{ij}G_{b_1b_2ij}
+c_2~R_{a_1a_2b_1b_2}   \right)\vert_{(02000)}\nn\\
B_{a_1\dots a_5,b_1b_2}&=l^3\left( c_3~\varepsilon_{a_1\dots a_5}{}^{ijklmn}
G_{ijkl}G_{b_1b_2mn} \right)\vert_{(01002)}\nn\\
C_{a_1\dots a_5,b_1\dots b_5}&=0~.
\la{marsalis}
\end{align}

Similarly for the fields $U$, $V$ we get
\begin{align}
V_{ab\al}&=l^3(c_4~T_{ab\al})\nn\\
U_{a_1\dots a_5\al}&=0
~,
\la{metheny}
\end{align}
where $c_1\dots c_4$ are arbitrary real constants at this stage.
For the constraints in (\ref{shorter}) we have
\begin{align}
I_A&=l^3\left(c_5~G_{a_1a_2}{}^{ij}(\ga_{b_1b_2}T_{ij})_\al 
\right)\vert_{(02001)}\nn\\
I_B&=0\nn\\
I_C&=0~,
\end{align}
where $c_5$ is a linear combination of $c_1$, $c_2$, $c_3$ which can
be determined from (\ref{shorter}), (\ref{marsalis}). 
The action of the spinorial derivative on 
the physical field-strengths is given in the appendix. 
We see that the $I_B$, $I_C$ constraints are automatically satisfied. 
Moreover, imposing $I_A=0$ fixes the ratio $c_1/c_3$ in terms 
of $c_2/c_3$.
Finally, using the freedom (\ref{brecker}) it is straightforward  
to see from (\ref{jarret}), (\ref{metheny}) 
 that the term in $A_{a_1a_2,b_1b_2}$ 
proportional to $R_{a_1a_2b_1b_2}\vert$ 
(cf. (\ref{marsalis})) can be redefined away for a suitable $c_4$. 
The above discussion is 
illustrated in figure 1.
\bigskip
\leavevmode
\begin{figure}[h] \label {order3}
\begin{center}
\includegraphics[scale=0.5]{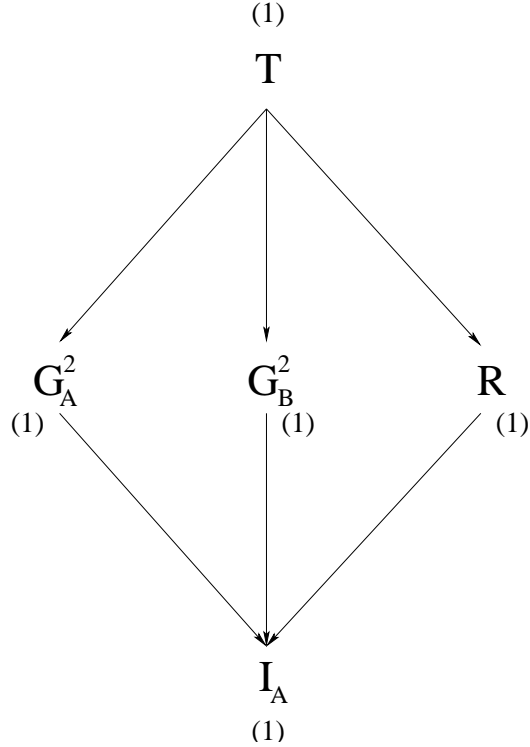}
\end{center}
\caption{Spinorial cohomology at  $l^3$. The first, 
second and third rows depict schematically 
all possible terms in $V$, $A\oplus B$  and  
$I_A$ respectively. The arrows indicate the action of $d_F$. 
Multiplicities are denoted by the numbers in parentheses.}
\end{figure}

To summarize, we have found that $H^{0,4}_{F}(phys)$
is generated by a certain linear combination
 of the two $G^2$ terms in (\ref{marsalis}).
This result can be understood alternatively as follows:
at order ${\cal O}(l^3)$ one can consider shifts of 
the superfour-form $G$ of the type 
\be
G\rightarrow G_\beta:=G-\frac{\beta l^3}{4\pi} trR^2~,
\la{corea}
\ee
where
\be
trR^2:=\frac{1}{4}E^DE^CE^BE^AR_{AB}{}^{ab}R_{CD ba}
\ee
and $\beta$ is an arbitrary real parameter.
Note that the top component of  $trR^2$ 
is proportional (taken at $\theta=0$) 
to the first Pontryagin class 
$p_1(M)$ of the eleven-dimensional 
spacetime manifold $M$, 
\be
p_1(M)=-\frac{1}{8\pi^2}trR^2~.
\ee
The lowest, purely spinorial, component 
of $trR^2$ is equal to a certain linear
combination of $G^2$ terms, as can be seen from 
(\ref{tara}) of the appendix. It could 
be brought to the form (\ref{rollins}) by a field
redefinition, but it will be 
convenient not
to do so here. If $G$ is closed so is $G_{\beta}$, by virtue of the second 
Bianchi identity in 
(\ref{parker}). Therefore, the resulting theory can be obtained from the
undeformed one simply by replacing $G$ with $G_\beta$ and this is the  
{\it only} resulting modification to the equations of motion. 
In other words, 
the generating element of $H^{0,4}_{F}(phys)$ 
can be absorbed by  a transformation of the form (\ref{corea}).

Locally (\ref{corea}) can be expressed as a 
shift in the superthree-form potential
\be
C\rightarrow C_\beta:=C-\frac{\beta l^3}{4\pi} Q~,
\la{spinoza}
\ee
where $Q$, $dQ=trR^2$, is the Chern-Simons form
\be
Q:=tr(\Om d\Om+\frac{2}{3}\Om^3)~.
\ee
Note however, that this is 
admissible as a field redefinition only in the
case $p_1(M)=0$. In the generic case, 
$p_1(M)\neq 0$, the cohomology
classes of $G$, $G_\beta$ are different, 
and the two theories related 
by the shift (\ref{spinoza}) are inequivalent.

At the level of classical actions, there is a one-parameter
family of supersymmetric theories given by \footnote{For simplicity
we concentrate on the bosonic part; we use the 
same letters ($C$, $G$, $R$) for the top components 
of the corresponding superforms. }
\begin{align}
S_{\beta}&=\frac{1}{l^9}\int\left(R{}^*{ 1} -\frac{1}{2}G_{\beta}
\wedge {}*G_{\beta} 
+\frac{1}{6}C_\beta\wedge G_{\beta}\wedge G_{\beta} +{\cal O}(l^6)
\right)_{\theta=0}~.
\la{one}
\end{align}
Note that unless one postulates an unconventional 
parity transformation law for the four-form $G$, 
(\ref{one}) is not parity-invariant. 
The actions (\ref{one}) are related to the ordinary 
supergravity action $S_0$ by a shift in the three-form potential
that changes the cohomology class of $G$:
\be
[\frac{G_\beta}{2\pi l^3}]=[\frac{G}{2\pi l^3}]+\beta p_1(M)~.
\label{gb}
\ee
On the other hand, quantum mechanical 
consistency of the theory forces  
$G$ to obey the quantization condition \cite{w}
\be
[\frac{G}{2\pi l^3}]-\frac{1}{4}p_1(M)\in H^4(M,\ZZ)~
\label{gq}
\ee
and therefore $G_\beta$ obeys a shifted  
quantization condition dictated by (\ref{gb}),(\ref{gq}).
%

\section*{Acknowledgments}

I am indebted to P.~Howe for valuable discussions. 
This work is supported by EU contract HPRN-CT-2000-00122.

\appendix
\section{Undeformed 11D supergravity in superspace}

The nonzero components of the supertorsion and supercurvature 
of undeformed 11D supergravity are given by
\begin{align} 
T_{\al\beta}{}^c&=-i(\ga^c)_{\al\beta}\nn\\
T_{a\beta}{}^{\ga}&= -{1\over36}\left((\ga^{bcd})_{\beta}{}^{\ga}
G_{abcd} +{1\over8} (\ga_{abcde})_{\beta}{}^{\ga} G^{abcd}\right)
\end{align}
and
\begin{align}
R_{\al\beta ab} &={i\over 6}\left((\ga^{cd})_{\al\beta} G_{abcd} +
{1\over24}(\ga_{abcdef})_{\al\beta} G^{cdef}\right)\nn\\ 
R_{\al bcd}&={i\over 2}\left( (\ga_bT_{cd})_\al+(\ga_cT_{bd})_\al
- (\ga_dT_{bc})_\al  \right)~.
\label{tara}
\end{align}
Note that the Lorentz condition implies
\be
R_{AB\al}{}^\beta=\frac{1}{4}R_{ABcd}(\ga^{cd})_\al{}^\beta~.
\ee
The action of the spinorial derivative on the physical
field strengths and their equations of motion are given by
\begin{align}
D_{\al}G_{abcd}&=6i(\ga_{[ab}T_{cd]})_\al\nn\\
D_{\al}T_{ab}{}^\beta&=\frac{1}{4}R_{abcd}(\ga^{cd})_{\al}{}^\beta
-2D_{[a}T_{b]\al}{}^\beta-2T_{[a|\al}{}^\epsilon T_{|b]\epsilon}{}^\beta\nn\\
D_{\al}R_{abcd}&=2D_{[a|}R_{\al|b]cd}-T_{ab}{}^\epsilon R_{\epsilon \al cd}
+2T_{[a|\al}{}^\epsilon R_{\epsilon|b]cd}
\end{align}
\la{derivatives}
and
\begin{align}
D_{[a}G_{bcde]}&=0\nn\\
D^fG_{fabc}&=-\frac{1}{2(4!)^2}
\varepsilon_{abcd_1\dots d_8}G^{d_1\dots d_4}G^{d_5\dots d_8} \nn\\
(\ga^aT_{ab})_\al&=0\nn\\
R_{ab}-\frac{1}{2}\eta_{ab}R&=-\frac{1}{12}\left(G_{adfg}G_b{}^{dfg}
-\frac{1}{8}\eta_{ab}G_{dfge}G^{dfge}  \right) ~.
\end{align}
The above equations can be integrated to an action whose
bosonic part reads
\be
S=\frac{1}{l^9}\int\left(R{}^*{ 1} -\frac{1}{2}G\wedge {}*G 
+\frac{1}{6}C\wedge G\wedge G\right)_{\theta=0}~.
\ee
%

%
%


\begin{thebibliography}{99}


\bibitem{duff}
M.~J.~Duff, ``The World In Eleven Dimensions: Supergravity,
Supermembranes And M-Theory,''
 Bristol, UK: IOP (1999).

\bibitem{t}
W.~Taylor,
``M(atrix) theory: Matrix quantum mechanics as a fundamental theory,''
Rev.\ Mod.\ Phys.\  {\bf 73} (2001) 419; hep-th/0101126.


\bibitem{cgnn}
M.~Cederwall, U.~Gran, M.~Nielsen and B.~E.~Nilsson, ``Manifestly
supersymmetric M-theory,'' hep-th/0007035, JHEP {\bf 0010} (2000) 041;
``Generalised 11-dimensional supergravity,'' hep-th/0010042.


\bibitem{pvw}
K.~Peeters, P.~Vanhove and A.~Westerberg, ``Supersymmetric
higher-derivative actions in ten and eleven dimensions,  the
associated superalgebras and their formulation in superspace,'' 
hep-th/0010167, Class.\ Quant.\ Grav.\  {\bf 18} (2001) 843.


\bibitem{pvwb}
K.~Peeters, P.~Vanhove and A.~Westerberg,
``Chiral splitting and world-sheet gravitinos in higher-derivative string
amplitudes,''
Class.\ Quant.\ Grav.\  {\bf 19} (2002) 2699; hep-th/0112157.


\bibitem{cnta}
M.~Cederwall, B.~E.~Nilsson and D.~Tsimpis, ``The structure of
maximally supersymmetric Yang-Mills theory: Constraining
higher-order corrections,'' JHEP {\bf 0106} (2001) 034; hep-th/0102009.



\bibitem{cntc}
M.~Cederwall, B.~E.~Nilsson and D.~Tsimpis, ``Spinorial cohomology
and maximally supersymmetric theories,'' JHEP {\bf 0202} (2002)
009; hep-th/0110069.

\bibitem{ht}
P.~S.~Howe and D.~Tsimpis,
``On higher-order corrections in M theory,''
JHEP {\bf 0309} (2003) 038; hep-th/0305129.


\bibitem{cntb}
M.~Cederwall, B.~E.~Nilsson and D.~Tsimpis, ``D = 10
super-Yang-Mills at $O(\alpha'^2)$,'' JHEP {\bf 0107} (2001) 042; 
hep-th/0104236.



\bibitem{cntd}
M.~Cederwall, B.~E.~Nilsson and D.~Tsimpis, ``Spinorial cohomology
of abelian d = 10 super-Yang-Mills at  $O(\alpha'^3)$,'' JHEP {\bf
0211} (2002) 023; hep-th/0205165.





\bibitem{hklt}
P.~S.~Howe, S.~F.~Kerstan, U.~Lindstrom and D.~Tsimpis,
``The deformed M2-brane,''
JHEP {\bf 0309} (2003) 013; hep-th/0307072.


\bibitem{kd}
J.~M.~Drummond and S.~F.~Kerstan,
``Kappa-symmetric derivative corrections to D-brane dynamics,'' 
hep-th/0407145.



\bibitem{hpa}
P.~S.~Howe, ``Pure Spinors Lines In Superspace And Ten-Dimensional
Supersymmetric Theories,'' Phys.\ Lett.\ B {\bf 258} (1991) 141
[Addendum-ibid.\ B {\bf 259} (1991) 511].

\bibitem{hpb}
P.~S.~Howe, ``Pure spinors, function superspaces and supergravity
theories in ten-dimensions and eleven-dimensions,'' Phys.\ Lett.\
B {\bf 273} (1991) 90.


\bibitem{ba}
N.~Berkovits,
``Super-Poincare covariant quantization of the superstring,''
JHEP {\bf 0004} (2000) 018; hep-th/0001035.



\bibitem{bb}
N.~Berkovits, ``Covariant quantization of the superparticle using
pure spinors,'' JHEP {\bf 0109} (2001) 016; hep-th/0105050.


\bibitem{gr}
P.~A.~Grassi, G.~Policastro, M.~Porrati and P.~Van Nieuwenhuizen,
``Covariant quantization of superstrings without pure spinor constraints,''
JHEP {\bf 0210} (2002) 054; hep-th/0112162. 
P.~A.~Grassi, G.~Policastro and P.~van Nieuwenhuizen,
``The massless spectrum of covariant superstrings,''
JHEP {\bf 0211} (2002) 001; hep-th/0202123.




\bibitem{w}
E.~Witten,
``On flux quantization in M-theory and the effective action,''
J.\ Geom.\ Phys.\  {\bf 22} (1997) 1; hep-th/9609122.

\bibitem{1}
D.~E.~Diaconescu, G.~W.~Moore and E.~Witten,
``E(8) gauge theory, and a derivation of K-theory from M-theory,''
Adv.\ Theor.\ Math.\ Phys.\  {\bf 6} (2003) 1031; hep-th/0005090.

\bibitem{2}
D.~E.~Diaconescu, G.~W.~Moore and E.~Witten,
``A derivation of K-theory from M-theory,'' hep-th/0005091.

\bibitem{3}
G.~W.~Moore and E.~Witten,
``Self-duality, Ramond-Ramond fields, and K-theory,''
JHEP {\bf 0005} (2000) 032; hep-th/9912279.

\bibitem{4}
G.~W.~Moore,
``Some comments on branes, G-flux, and K-theory,''
Int.\ J.\ Mod.\ Phys.\ A {\bf 16} (2001) 936; hep-th/0012007.

\bibitem{5}
E.~Diaconescu, G.~Moore and D.~S.~Freed,
``The M-theory 3-form and E(8) gauge theory,'' hep-th/0312069.


\bibitem{bl}
L.~Bonora, K.~Lechner, M.~Bregola, P.~Pasti and M.~Tonin, ``A
Discussion Of The Constraints In N=1 Sugra-Sym In 10-D,'' Int.\
J.\ Mod.\ Phys.\ A {\bf 5} (1990) 461.





\bibitem{cjs}
E.~Cremmer, B.~Julia and J.~Scherk, ``Supergravity Theory In 11
Dimensions,'' Phys.\ Lett.\ B {\bf 76} (1978) 409.


\bibitem{cf}
E.~Cremmer and S.~Ferrara, ``Formulation Of Eleven-Dimensional
Supergravity In Superspace,'' Phys.\ Lett.\ B {\bf 91} (1980) 61.


\bibitem{brh}
L.~Brink and P.~S.~Howe, ``Eleven-Dimensional Supergravity On The
Mass - Shell In Superspace,'' Phys.\ Lett.\ B {\bf 91} (1980) 384.


\bibitem{dragon}
N.~Dragon, ``Torsion And Curvature In Extended Supergravity,'' Z.\
Phys.\ C {\bf 2} (1979) 29.


\bibitem{hw}
P.~S.~Howe, ``Weyl superspace,'' hep-th/9707184, 
Phys.\ Lett.\ B {\bf 415} (1997) 149.

\bibitem{nr}
H.~Nishino and S.~Rajpoot, ``A note on embedding of M-theory
corrections into eleven-dimensional  superspace,'' Phys.\ Rev.\ D
{\bf 64} (2001) 124016; hep-th/0103224.








\end{thebibliography}
\end{document}